\documentstyle[twoside, epsfig]{article}

\catcode`\@=11
\long\def\@makefntext#1{
\protect\noindent \hbox to 3.2pt {\hskip-.9pt
$^{{\eightrm\@thefnmark}}$\hfil}#1\hfill}               %CAN BE USED

\def\@makefnmark{\hbox to 0pt{$^{\@thefnmark}$\hss}}    %ORIGINAL

\def\ps@myheadings{\let\@mkboth\@gobbletwo
\def\@oddhead{\hbox{}
\rightmark\hfil\eightrm\thepage}
\def\@oddfoot{}\def\@evenhead{\eightrm\thepage\hfil
\leftmark\hbox{}}\def\@evenfoot{}
\def\sectionmark##1{}\def\subsectionmark##1{}}

\oddsidemargin=\evensidemargin
\newcounter{sectionc}\newcounter{subsectionc}\newcounter{subsubsectionc}
\renewcommand{\section}[1] {\vspace{12pt}\addtocounter{sectionc}{1}
\setcounter{subsectionc}{0}\setcounter{subsubsectionc}{0}\noindent
        {\tenbf\thesectionc. #1}\par\vspace{5pt}}
\renewcommand{\subsection}[1] {\vspace{12pt}\addtocounter{subsectionc}{1}
      \setcounter{subsubsectionc}{0}\noindent
      {\bf\thesectionc.\thesubsectionc.{\kern1pt \bfit #1}}\par\vspace{5pt}}
\renewcommand{\subsubsection}[1]
      {\vspace{12pt}\addtocounter{subsubsectionc}{1}
      \noindent{\tenrm\thesectionc.\thesubsectionc.\thesubsubsectionc.
      {\kern1pt \tenit #1}}\par\vspace{5pt}}
\newcommand{\nonumsection}[1] {\vspace{12pt}\noindent{\tenbf #1}
        \par\vspace{5pt}}

\newcounter{appendixc}
\newcounter{subappendixc}[appendixc]
\newcounter{subsubappendixc}[subappendixc]
\renewcommand{\thesubappendixc}{\Alph{appendixc}.\arabic{subappendixc}}
\renewcommand{\thesubsubappendixc}
        {\Alph{appendixc}.\arabic{subappendixc}.\arabic{subsubappendixc}}

\renewcommand{\appendix}[1] {\vspace{12pt}
        \refstepcounter{appendixc}
        \setcounter{figure}{0}
        \setcounter{table}{0}
        \setcounter{lemma}{0}
        \setcounter{theorem}{0}
        \setcounter{corollary}{0}
        \setcounter{definition}{0}
        \setcounter{equation}{0}
        \renewcommand{\thefigure}{\Alph{appendixc}.\arabic{figure}}
        \renewcommand{\thetable}{\Alph{appendixc}.\arabic{table}}
        \renewcommand{\theappendixc}{\Alph{appendixc}}
        \renewcommand{\thelemma}{\Alph{appendixc}.\arabic{lemma}}
        \renewcommand{\thetheorem}{\Alph{appendixc}.\arabic{theorem}}
        \renewcommand{\thedefinition}{\Alph{appendixc}.\arabic{definition}}
        \renewcommand{\thecorollary}{\Alph{appendixc}.\arabic{corollary}}
        \renewcommand{\theequation}{\Alph{appendixc}.\arabic{equation}}
        \noindent{\tenbf Appendix \theappendixc #1}\par\vspace{5pt}}
\newcommand{\subappendix}[1] {\vspace{12pt}
        \refstepcounter{subappendixc}
        \noindent{\bf Appendix \thesubappendixc. {\kern1pt \bfit #1}}
        \par\vspace{5pt}}
\newcommand{\subsubappendix}[1] {\vspace{12pt}
        \refstepcounter{subsubappendixc}
        \noindent{\rm Appendix \thesubsubappendixc. {\kern1pt \tenit #1}}
        \par\vspace{5pt}}

\newcommand{\smalllineskip}{\baselineskip=10pt}

\def\eightcirc{
\begin{picture}(0,0)
\put(4.4,1.8){\circle{6.5}}
\end{picture}}
\def\eightcopyright{\eightcirc\kern2.7pt\hbox{\eightrm c}}

\def\abstracts#1#2#3{{
        \centering{\begin{minipage}{4.5in}\baselineskip=10pt\footnotesize
        \parindent=0pt #1\par
        \parindent=15pt #2\par
        \parindent=15pt #3
        \end{minipage}}\par}}

\renewenvironment{thebibliography}[1]
        {\frenchspacing
         \ninerm\baselineskip=11pt
         \begin{list}{\arabic{enumi}.}
        {\usecounter{enumi}\setlength{\parsep}{0pt}
         \setlength{\leftmargin 12.7pt}{\rightmargin 0pt} %FOR 1--9 ITEMS
         \setlength{\itemsep}{0pt} \settowidth
        {\labelwidth}{#1.}\sloppy}}{\end{list}}

\newcounter{itemlistc}
\newcounter{romanlistc}
\newcounter{alphlistc}
\newcounter{arabiclistc}

\newcommand{\fcaption}[1]{
        \refstepcounter{figure}
        \setbox\@tempboxa = \hbox{\footnotesize Fig.~\thefigure. #1}
        \ifdim \wd\@tempboxa > 5in
           {\begin{center}
        \parbox{5in}{\footnotesize\smalllineskip Fig.~\thefigure. #1}
            \end{center}}
        \else
             {\begin{center}
             {\footnotesize Fig.~\thefigure. #1}
              \end{center}}
        \fi}

\newcommand{\tcaption}[1]{
        \refstepcounter{table}
        \setbox\@tempboxa = \hbox{\footnotesize Table~\thetable. #1}
        \ifdim \wd\@tempboxa > 5in
           {\begin{center}
        \parbox{5in}{\footnotesize\smalllineskip Table~\thetable. #1}
            \end{center}}
        \else
             {\begin{center}
             {\footnotesize Table~\thetable. #1}
              \end{center}}
        \fi}

\def\@citex[#1]#2{\if@filesw\immediate\write\@auxout
        {\string\citation{#2}}\fi
\def\@citea{}\@cite{\@for\@citeb:=#2\do
        {\@citea\def\@citea{,}\@ifundefined
        {b@\@citeb}{{\bf ?}\@warning
        {Citation `\@citeb' on page \thepage \space undefined}}
        {\csname b@\@citeb\endcsname}}}{#1}}

\newif\if@cghi
\def\cite{\@cghitrue\@ifnextchar [{\@tempswatrue
        \@citex}{\@tempswafalse\@citex[]}}
\def\citelow{\@cghifalse\@ifnextchar [{\@tempswatrue
        \@citex}{\@tempswafalse\@citex[]}}
\def\@cite#1#2{{$\null^{#1}$\if@tempswa\typeout
        {IJCGA warning: optional citation argument
        ignored: `#2'} \fi}}

\def\@refcitex[#1]#2{\if@filesw\immediate\write\@auxout
        {\string\citation{#2}}\fi
\def\@citea{}\@refcite{\@for\@citeb:=#2\do
        {\@citea\def\@citea{, }\@ifundefined
        {b@\@citeb}{{\bf ?}\@warning
        {Citation `\@citeb' on page \thepage \space undefined}}
        \hbox{\csname b@\@citeb\endcsname}}}{#1}}

\def\@refcite#1#2{{#1\if@tempswa\typeout
        {IJCGA warning: optional citation argument
        ignored: `#2'} \fi}}

\def\refcite{\@ifnextchar[{\@tempswatrue
        \@refcitex}{\@tempswafalse\@refcitex[]}}

\def\pmb#1{\setbox0=\hbox{#1}
        \kern-.025em\copy0\kern-\wd0
        \kern.05em\copy0\kern-\wd0
        \kern-.025em\raise.0433em\box0}

\def\fnt#1#2{\footnotetext{\kern-.3em
        {$^{\mbox{\scriptsize #1}}$}{#2}}}

\def\fpage#1{\begingroup
\voffset=.3in
\thispagestyle{empty}\begin{table}[b]\centerline{\footnotesize #1}
        \end{table}\endgroup}

\def\runninghead#1#2{\pagestyle{myheadings}
\markboth{{\protect\footnotesize\it{\quad #1}}\hfill}
{\hfill{\protect\footnotesize\it{#2\quad}}}}
\font\tenrm=cmr10
\font\tenit=cmti10
\font\tenbf=cmbx10
\font\bfit=cmbxti10 at 10pt
\font\ninerm=cmr9

\font\eightrm=cmr8

\def\qed{\hbox{${\vcenter{\vbox{                      %HOLLOW SQUARE
   \hrule height 0.4pt\hbox{\vrule width 0.4pt height 6pt
   \kern5pt\vrule width 0.4pt}\hrule height 0.4pt}}}$}}

\begin{document}

\runninghead {E. Comay}
{A Regular Theory of Magnetic $\ldots$}

%\normalsize\textlineskip
\thispagestyle{empty}\setcounter{page}{82}
\vspace*{0.88truein}
\fpage{82}

\centerline{\bf A REGULAR THEORY OF MAGNETIC MONOPOLES}
\vspace*{0.035truein}
\centerline{\bf  AND ITS IMPLICATIONS}
\vspace*{0.035truein}

\vspace*{0.37truein}
\centerline{\footnotesize E. Comay}

\centerline{\footnotesize \it
 School of Physics and Astronomy Raymond and Beverly Sackler Faculty of
Exact Sciences}
\baselineskip=10pt
\centerline{\footnotesize \it Tel Aviv
University, Tel Aviv 69978, Israel}

%\date{\today}

\baselineskip 5mm

\vspace*{0.21truein}

\abstracts{A  regular charge-monopole theory is derived from simple and
self-evident postulates.  It is shown that this theory provides
explanations for effects  of strong and nuclear interactions. The theory
is compared with Dirac's  monopole theory. Applications to strong and
nuclear interactions are  compared with quantum chromodynamics. The
results favor the regular  charge-monopole theory and indicate
difficulties of the other ones.  An experiment that may provide further
evidence  helping to decide between  the regular charge-monopole theory
and quantum chromodynamics is suggested.}{}{}

%\pacs{PACS numbers: 0.0, 0.0.}

\bigskip

$$$$

\section{Introduction}

Classical electrodynamics is regarded as a well established part  of
physics.  The equations of Maxwell and the Lorentz law of force are the equations
of motion of a system of electromagnetic fields and electrically charged
matter. The fields part of these equations consists of two kinds of
entities - namely, the electric and the magnetic fields. On the other
hand, the massive matter part contains a single kind of electromagnetic
entity - the electric charge.

The idea of  magnetic monopoles (called just monopoles in the literature)
aims to restore the symmetry between electricity and magnetism. Let us
examine the electromagnetic fields tensor[1,2]

\begin{equation}
F_{(e)}^{\mu \nu } = \left(
\begin{array}{cccc}
0   & -E_x & -E_y & -E_z \\

E_x &  0   & -B_z &  B_y \\

E_y &  B_z &   0  & -B_x \\

E_z & -B_y &  B_x &  0
\end{array}
\right).
\label{eq:FE}
\end{equation}

In this work subscripts $_{(e)}$ and $_{(m)}$  denote  quantities related
to charges and monopoles, respectively. Units where the speed of light
$c=1$ are used. The Lorentz metric is diagonal and its entries are
$(1,-1,-1,-1)$. Greek indices run from 0 to 3. The symbol $_{,\mu }$
denotes the partial differentiation with respect to $x^\mu $. ${\bf i}$,
${\bf j}$ and ${\bf k}$ denote unit vectors in the $x$, $y$ and $z$
directions, respectively.

Duality transformations  cast a system of fields and charges into a system
of fields and monopoles. These transformations are (see [2], pp. 252, 551)

\begin{equation}
\bf E \rightarrow \bf B,\;\;\;\bf B \rightarrow -\bf E
\label{eq:EB}
\end{equation}

and

\begin{equation}
e \rightarrow g,\;\;\;g\rightarrow -e,
\label{eq:EG}
\end{equation}
where $g$ denotes the monopole strength. (In this work, duality is used in
its restricted form of a duality rotation by $\pi/2$.) Eq.
$(\!\!~\ref{eq:EG})$ represents relations between Lorentz scalars (see a
discussion of this issue in Section 4) whereas $(\!\!~\ref{eq:EB})$ takes
a relativistic tensorial form

\begin{equation}
F_{(m)}^{*\mu \nu } = \frac {1}{2}\varepsilon ^{\mu \nu \alpha \beta }
F_{(e)\alpha \beta }.
\label{eq:DUE}
\end{equation}
Here $\varepsilon ^{\mu \nu \alpha \beta }$ is the completely
antisymmetric unit tensor of the fourth rank. Entries of the monopole
dependent fields tensors of $(\!\!~\ref{eq:DUE})$ are

\begin{equation}
F_{(m)}^{*\mu \nu } = \left(
\begin{array}{cccc}
0   & -B_x & -B_y & -B_z \\
B_x &  0   &  E_z & -E_y \\
B_y & -E_z &   0  &  E_x \\
B_z &  E_y & -E_x &  0
\end{array}
\right).
\label{eq:FM}
\end{equation}

Historically, classical electrodynamics of charges and fields has been
formulated after a lot of experimental data have been accumulated. Results
of these experiments have guided physicists how to construct the theory.
Thus, in the case of electrodynamics of charges and fields, development
has taken the inductive way. (The addition of the Maxwell term makes an
exception.) Unfortunately, monopoles have not been detected in
laboratories. Hence, one cannot follow the historical course of classical
electrodynamics of charges and fields. Therefore, the deductive course
must be adopted and the theory should be derive from an appropriate set of
postulates.

Consider two postulates pertaining to this topic:

\begin{itemize}

\item[{(A)}] The theory of monopoles and fields takes a form which is
completely dual to the theory of charges and fields.

\item[{(B)}] Electromagnetic fields of a system of monopoles and those of
a system of charges have identical dynamical properties.

\end{itemize}

Hereafter, these postulates are called postulate (A) and (B),
respectively.  Upper case letters denote postulates and numbers denote
other kinds of items.

One may be tempted to use both postulates (A) and (B) as fundamental
elements of the theory. However, it turns out that this goal is
unattainable because different sets of equations of motion are obtained
from postulate (A) (without (B)) and from postulate (B) (without (A)).
Section 4 contains a discussion of this topic.

In Section 2, a regular charge-monopole theory is derived from postulate
(A). The relevance of this theory to strongly interacting particles (called
hadrons) and to nuclear interactions is discussed in Section 3. A
comparison of results obtained in Sections 2 with the Dirac monopole
theory (which is consistent with postulate (B)) is presented in Section 4.
A comparison of the results of Section 3 with quantum chromodynamics (QCD)
is shown in Section 5. Concluding remarks are the contents of the last
Section.

\section{A Regular Charge-Monopole Theory}

Postulate (A) means that classical electrodynamics of charges and their
associated fields is a cornerstone of the charge-monopole theory developed
here. For this reason, let us examine the equations of motion of classical
electrodynamics of charges and fields as well as some of its fundamental
relations. The equations of motion of the fields are Maxwell equations.
Their tensorial form is (see [1], pp. 74, 67 or [2], pp. 551)

\begin{equation}
F^{\mu \nu}_{(e)\;,\nu} = -4\pi j_{(e)}^\mu
\label{eq:MAX1}
\end{equation}

and

\begin{equation}
F^{*\mu \nu}_{(e)\;,\nu} = 0.
\label{eq:MAX2}
\end{equation}
where $F_{(e)}^{*\mu \nu}$ takes the same form as $(\!\!~\ref{eq:FM})$ but
here it represents fields of charges. The Lorentz law of force is the
equation of motion of charges (see [2], p. 572)

\begin{equation}
ma_{(e)}^\mu = eF_{(e)}^{\mu \nu }v_\nu,
\label{eq:LOR}
\end{equation}
where $v^\mu $ denotes the particle's 4-velocity, $m$ is its rest mass
and $e$ is its charge.

The fields used in $(\!\!~\ref{eq:MAX1})$ and in $(\!\!~\ref{eq:MAX2})$
can be derived from a regular 4-potential

\begin{equation}
F_{(e)\mu \nu } = A_{(e)\nu ,\mu } - A_{(e)\mu ,\nu }.
\label{eq:POT}
\end{equation}
(Point infinities associated with the introduction of elementary classical
point charges are beyond the scope of this work. A regular solution of
point infinity problems is presented in [3].)

The 4-potential is an important quantity of the system's dynamics, because
it is used in the charge-fields interaction term of the Lagrangian density
(see [1], p. 71 or [2], p. 596)

\begin{equation}
L_{int} = -j_{(e)}^\mu A_{(e)\mu}.
\label{eq:LINT}
\end{equation}

The reader should note the difference between the inhomogeneous Maxwell
equations $(\!\!~\ref{eq:MAX1})$ and the homogeneous ones
$(\!\!~\ref{eq:MAX2})$.  As a matter of fact, the homogeneous equation
$(\!\!~\ref{eq:MAX2})$ represents internal mathematical relations between
field components (see [1], pp. 66,67). On the other hand, the
inhomogeneous Maxwell equations $(\!\!~\ref{eq:MAX1})$ are associated with
charge-fields interaction and are derived from the Lagrangian density (see
[1], pp. 73-75 or [2], pp 597).

Let us now use postulate (A) and obtain the equations of motion of a
system of monopoles and their associated fields (namely, a system that
does not contain electric charges). This goal is achieved from the
application of the duality relations $(\!\!~\ref{eq:EB})$ -
$(\!\!~\ref{eq:DUE})$ to $(\!\!~\ref{eq:MAX1})$ - $(\!\!~\ref{eq:LOR})$.
The results are

\begin{equation}
F^{*\mu \nu}_{(m)\;,\nu} = -4\pi j_{(m)}^\mu,
\label{eq:MAX1M}
\end{equation}

\begin{equation}
-F^{\mu \nu}_{(m)\;,\nu} = 0.
\label{eq:MAX2M}
\end{equation}

and

\begin{equation}
ma_{(m)}^\mu = gF_{(m)}^{*\mu \nu }v_\nu.
\label{eq:LORM}
\end{equation}

One should note that $F_{(m)}^{*\mu \nu}$ of $(\!\!~\ref{eq:MAX1M})$ takes
the form of $(\!\!~\ref{eq:FM})$ and $F_{(m)}^{\mu \nu}$ of
$(\!\!~\ref{eq:MAX2M})$ takes the form of $(\!\!~\ref{eq:FE})$. Like in
the case of charges, the fields of monopoles can be derived from an
appropriate 4-potential

\begin{equation}
F^*_{(m)\mu \nu } = A_{(m)\nu ,\mu } - A_{(m)\mu ,\nu }.
\label{eq:POTM}
\end{equation}
For this system, the interaction part of the Lagrangian density is
analogous to $(\!\!~\ref{eq:LINT})$

\begin{equation}
L_{int} = -j_{(m)}^\mu A_{(m)\mu}.
\label{eq:LINTM}
\end{equation}

The reader should note that in the monopole case, the fields tensors take
the opposite role, with respect to that of charges. Thus, $F^{*\mu\nu}_{(m)}$ of $(\!\!~\ref{eq:MAX1M})$ is related to the interaction term
$(\!\!~\ref{eq:LINTM})$ whereas $F_{(m)}^{\mu \nu }$ of
$(\!\!~\ref{eq:MAX2M})$ represents mathematical relations.
At this point we have the equations of motion of two noninteracting
systems: a system of charges and their associated fields and another
system which consists of monopoles and their associated fields. The rest
of this Section is devoted to the equations of motion of a combined system
containing charges, monopoles and their fields.

Due to the linearity of electrodynamics, one may split the electromagnetic
fields into a sum of field quantities and examine their effects
separately. An important kind of split is the one which examines bound
fields and radiation fields separately. (A decomposition of the
electromagnetic fields of a system of charges into these components can be
found in [3].) The first kind of fields is significant near the charges
and decays rapidly at larger distances. Radiation fields decay slower at
larger distances and become dominant there. They represent an entity which
is distinguished from charges. In classical physics they take the form of
radiation energy and in quantum mechanics they appear as real photons.

Radiation fields emitted  from a specific source have the following
properties

\begin{equation}
B^2 - E^2 = 0
\label{eq:EEQB}
\end{equation}

and

\begin{equation}
{\bf E\cdot B} = 0.
\label{eq:EDOTB}
\end{equation}
(These relations are obtained from eq. (66.8) of [1], p. 172 and from eqs.
(9.4) and (9.5) of [2], p. 392.) Relations $(\!\!~\ref{eq:EEQB})$ and
$(\!\!~\ref{eq:EDOTB})$ are just a reformulation of the Lorentz scalar
$F^{\mu \nu}F_{\mu \nu}$ and the pseudoscalar $F^{*\mu \nu}F_{\mu \nu}$,
respectively. An observation of these quantities proves that in the case
of radiation fields, they remain invariant under the duality
transformation  $(\!\!~\ref{eq:EB})$. This result indicates that radiation
fields of charges and radiation fields of monopoles may be regarded as one
and the same entity. This result is also obtained from  the equations of
motion of radiation fields.  These equations are the {\em homogeneous}
Maxwell equations

\begin{equation}
F^{\mu \nu}_{(e)\;,\nu} = 0
\label{eq:MAXH1}
\end{equation}

and

\begin{equation}
F^{*\mu \nu}_{(e)\;,\nu} = 0.
\label{eq:MAXH2}
\end{equation}

Let us examine the simple example of linearly polarized monochromatic
plane wave (see [1], pp. 114-117 or [2], pp. 273-278) which propagates in
the $z$-direction. Taking the electric charge point of view, a vector
potential of the fields is

\begin{equation}
{\bf A}_{(e)} = -iAe^{i\omega (z-t)}{\bf i}.
\label{eq:APW}
\end{equation}

Thus, we have

\begin{equation}
{\bf E} = -\frac {\partial A_{(e)}}{\partial t} =
\omega Ae^{i\omega (z-t)}{\bf i}
\label{eq:EPW}
\end{equation}

and

\begin{equation}
{\bf B} = \nabla \times A_{(e)} = \omega Ae^{i\omega (z-t)}{\bf j.}
\label{eq:BPW}
\end{equation}
(Obviously, the real part of these equations represents the quantities
that should be accounted for.) The direction of the linear polarization is
that of the electric field $(\!\!~\ref{eq:EPW})$.

Let us now take the monopole point of view. Here one  examines a vector
potential which is parallel to the y-axis

\begin{equation}
{\bf A}_{(m)} = -iAe^{i\omega (z-t)}{\bf j}.
\label{eq:APWM}
\end{equation}

Using the duality relations $(\!\!~\ref{eq:EB})$ as well as
$(\!\!~\ref{eq:POTM})$, one finds

\begin{equation}
{\bf B} = -\frac {\partial A_{(m)}}{\partial t} =
\omega Ae^{i\omega (z-t)}{\bf j}
\label{eq:BPWM}
\end{equation}

and

\begin{equation}
{\bf E} = -\nabla \times A_{(m)} = \omega Ae^{i\omega (z-t)}{\bf i}.
\label{eq:EPWM}
\end{equation}

It follows that $(\!\!~\ref{eq:EPW})$ equals $(\!\!~\ref{eq:EPWM})$ and
$(\!\!~\ref{eq:BPW})$ equals $(\!\!~\ref{eq:BPWM})$.

These results indicate explicitly that one may identify radiation fields
of charges with radiation fields of monopoles without arriving at any
contradiction. Henceforth, radiation fields of charges and radiation
fields of monopoles are regarded as one and the same entity and are
denoted  by the subscript $_{(w)}$. Thus, for example, 
$F^{\mu\nu}_{(e,w)}$ denotes the tensor of bound and radiation fields of 
charges and the radiation fields of monopoles. The symbol $F^{\mu 
\nu}_{(m,w)}$ is analogous.

Let us examine bound fields of charges and bound fields of monopoles. A
simple example is the system which consists of one charge and one
monopole. The interaction term of the Lagrangian density is required.
Again, due to the linearity of electrodynamics, one may write the
interaction part of a system which consists of many charges and many
monopoles as a sum of two body interactions. There are several
restrictions imposed on the form of the required quantity:

\begin{itemize}

\item[{I.}] Since the action is a Lorentz scalar, all terms of the
Lagrangian density must be Lorentz scalars too.

\item[{II.}] Due to the linearity of electrodynamics, the charge-monopole
interaction term must be a sum of bilinear quantities containing two
factors, one is related to charges and the other is related to monopoles.

\item[{III.}] Due to the notion of a charge and of a monopole, a system of
one motionless charge and one motionless monopole does not change in time.

\end{itemize}

The charge-charge interaction term $(\!\!~\ref{eq:LINT})$ satisfies
requirements I and the appropriate analogue of II. Relation
$(\!\!~\ref{eq:LINTM})$ is the monopole version of $(\!\!~\ref{eq:LINT})$.
An observation of a system of one motionless charge and one motionless
monopole proves that, unlike the case of radiation fields, bound fields of
these objects differ substantially. Thus, the Lorentz scalar
$(\!\!~\ref{eq:EEQB})$ $B^2 -  E^2 <0$ for the case of a charge and $B^2 
-E^2 >0$ for the case of a monopole.

A simple analysis[4] proves that in the case of bound fields, one cannot
create an interaction term that satisfies I, II and III as well as some
other self-evident postulates. Thus, the following conclusion is obtained:

\begin{itemize}

\item[{1.}] Charges do not interact with bound fields of monopoles and
monopoles do not interact with bound fields of charges. Charges interact
with all fields of charges and with radiation fields emitted from
monopoles. Monopoles interact with all fields of monopoles and with
radiation fields emitted from charges.

\end{itemize}

Henceforth, this conclusion is referred to as conclusion 1. Conclusion 1
yields the following form of the Lorentz force exerted on charges

\begin{equation}
ma^\mu _{(e)} = eF_{(e,w)}^{\mu \nu }v_{(e)\nu}.
\label{eq:FINALLORE}
\end{equation}

The corresponding force exerted on monopoles is

\begin{equation}
ma^\mu _{(m)} = gF_{(m,w)}^{*\mu \nu }v_{(m)\nu}.
\label{eq:FINALLORM}
\end{equation}

Explicit expressions of the Lagrangian of the system and of the
energy-momentum tensor can be found in [4]. Since fields of charges and
fields of monopoles have different dynamical properties, it is suggested
that fields of monopoles be called magnetoelectric fields[4].

It is interesting to note that conclusion 1 can be obtained even without
the requirement of having a regular interaction term of the
charge-monopole Lagrangian density. It can be shown[5] that the same
result is obtained if one uses the postulate that, like the electric
charge, the magnetic monopole is a Lorentz scalar and not a pseudoscalar
(see [2], p. 253).

Conclusion 1 is the main result of the theoretical analysis carried out in
this section. It explains why postulates (A) and (B) which are written in
the introduction, cannot be used together as elements of a charge-monopole
theory. Indeed, if postulate (B) is adopted (like in Dirac's
charge-monopole theory[6-8]) then the Lorentz force exerted on a charge
must be $eF_{(e,m,w)}^{\mu \nu }v_{(e)\nu }$, contrary to
$(\!\!~\ref{eq:FINALLORE})$. Physical implications of conclusion 1 are
discussed in the next section.  In what follows, the monopole theory
outlined here is called the regular charge-monopole theory.

\clearpage

\section{A Discussion of Experimental Data}

Consider the 4 different kinds of interactions known in physics:
strong, electromagnetic, weak and gravitational. The gravitational
interaction is practically blind to many features of matter and
``sees" only the matter's energy-momentum tensor and that of
electromagnetic fields. For this reason, it is not mentioned here any
more. Table 1 shows 2 conservation properties of the other interactions,

\begin{table}[h]
\begin{center}
\begin{tabular}{|c|c|c|c|}
\multicolumn{4}{c}{} \\ \hline

& strong & electromagnetic & weak  \\ \hline

parity & yes    & yes             & no    \\ \hline

flavor & yes    & yes             & no    \\ \hline
\end{tabular}

\vglue 0.08in

{\caption{}

Validity of parity and flavor conservation \\
under three kinds of interactions}
\end{center}
\end{table}

The meaning of parity conservation is that if, for example, a theory of
the specific kind of force contains a Hamiltonian then this Hamiltonian
commutes with the parity operator. Parity conservation indicates
similarity between strong and electromagnetic interactions.
Flavor is a property of elementary constituents of matter: each one of the
3 kinds of charged leptons (electrons, muons and tau mesons) and each one
of the 6 kinds of quarks has its own flavor. Hereafter, the symbol $q$ is
sometimes used as a notation of a quark. All these objects are spin 1/2
Dirac particles and each one of them has its own antiparticle (generally,
an antiparticle is denoted by a bar over the particle's symbol). Table 1
shows that strong and electromagnetic interactions cannot alter flavor
properties of matter. Moreover, experiments show that these interactions
``see" quarks and leptons  as very small objects whose radius is less than
$10^{-16}cm$. As a matter of fact, it is a common belief that leptons and
quarks are pointlike (see [9], pp. 264, 276, 277). Experiments also show
that strong and electromagnetic interactions can perform a process called
pair production. This process results in the creation of a particle and an
antiparticle of the same kind, thereby obeying flavor conservation. Pair
production is seen in the creation of separate charged leptons, like an
electron and a positron or, for example, in the creation of mesons which
are $q\bar{q}$ bound systems. Table 1 indicates that strong and
electromagnetic interactions are similar with respect to flavor
conservation too.

On the basis of these results it is postulated here that

\begin{itemize}

\item[{(C)}] Strong interactions are interactions between magnetic
monopoles.  Thus, all charged leptons have a unit of electric charge $\pm 
e$ and no magnetic monopole, whereas quarks have both electric charge and
a magnetic monopole unit $-g$.

\end{itemize}

Applications of the regular charge-monopole theory and of postulate (C) to
experimental data are discussed briefly in this Section. As shown below,
several fundamental properties of the relevant data can be explained in a
qualitative manner. This point can be put in a different way. Postulate
(C) can be {\em refuted} if it fails to explain qualitative properties of
experimental data.

It is obvious that a classical theory cannot go far in explaining
properties of matter. Thus, additional postulates are used below, where
required. These postulates look natural and rely on fundamental properties
of matter. Putting them in the status of postulates does not negate the
possibility of proving them by means of a more profound theory. Up to now,
only postulates (A) and (C) are used.

\begin{itemize}

\item[{1.}] The regular charge-monopole theory of Section 2 predicts that
electromagnetic-like forces come in pairs, one is charge related and the
other pertains to monopoles. If, for example, 3 different kinds of such
interactions are found in Nature then the regular monopole theory is
refuted (or has to find rescue in an additional postulate which is
analogous to the one stating that ``monopoles do not exist").

\end{itemize}

Up to now, nothing is said on specific properties of magnetic charges. The
following postulate fills this gap.

\begin{itemize}
\item[{(D)}] Like the case of electric charge, the elementary unit $g$ of
magnetic charge is quantized. Moreover, the size of $g$ is rather big $g^2\gg e^2\simeq 1/137$. Experimental data indicate that $g^2\simeq 1$ (see
[9], p. 20).

\end{itemize}

The reader should note that in Dirac's charge-monopole theory which is
derived in accordance with postulate (B) of the introduction, there is a
numerical relation between the size of the elementary electric charge and
of its magnetic counterpart and $g^2 = 137/4$[6-8]. This relation, which
imposes a restriction on the size of the elementary magnetic charge, does
not hold in the regular charge-monopole theory derived in Section 2. Here
the size of the elementary magnetic unit is a free parameter.

Let us proceed further, using postulate (D).

\begin{itemize}

\item[{2.}]

Experiments show that elementary particles that have just electric
charges, like the electron, do not participate in strong interactions[10].
Moreover, the electric charge of proton's quarks is not identical to the
corresponding quantity of the neutron.  It turns out that energetic
electrons interact {\em differently} with quark constituents of protons
and neutrons (see [10], p. 200 and [11]). On the other hand, energetic
real photons interact strongly with quark targets of protons and neutrons
and, in these interactions, protons and neutrons look very much alike[12].
These properties of Nature fit like a glove Conclusion 1 of the regular
charge-monopole theory outlined in Section 2. Indeed, one has just to
replace the terms 'electric charge' by 'electron', 'magnetic charge' by
'quark' and use Postulate (D) which means that the monopole charge of
quarks dominates the process. The reader should note that Conclusion 1 is
obtained from a pure theoretical analysis that relies on simple postulates
which have (at least apparently) no relevance to the experimental data
mentioned here. Further aspects of this issue are discussed in Section 5.

\end{itemize}

In order to proceed further, we need the following postulate, which relies
on well established experimental data (see [9], pp. 275-277)

\begin{itemize}

\item[{(E)}] Quarks are spin 1/2 Dirac particles.

\end{itemize}

This postulate enables us to make further deductions.

\begin{itemize}

\item[{3.}] Theory predicts and experiments show bound states of an
electron and a positron, called positronium. Now, by the postulates which
we already have at hand, one predicts strongly bound $q\bar {q}$ states
whose total spin and parity are analogous to the states of the
positronium. The fact that there are 6 different kinds of quarks, all of
which are assumed to have the same elementary magnetic charge, indicates
that mesonic states are much richer than the positronium ones, because the
quarks of a $q\bar {q}$ pair may or may not have the same type of flavor.
This prediction is supported by experimental data of mesons[13].

\item[{4.}] Mesons whose quantum states cannot be created by a $q\bar {q}$
system are called exotic. Binding forces between 2 mesons, namely a
$qq\bar q \bar q$ system, are expected to be much weaker than the $q\bar{q}$ bond. This conclusion is analogous to the relation between the value
of the binding energy of electrons in atoms and that of the molecular (or
van der Waals) forces between neutral atoms (and neutral molecules). In
particular, the lowest $q\bar {q}$ state (called $\pi $ meson) has a total
spin=0 and is analogous to an atom of a noble gas. Thus, one expects that
if bound states of a $\pi $ meson and another hadron exist at all then
their binding energy must be very small. Also in this case, predictions
are supported by experimental data[13].

\end{itemize}

A set of particles called baryons is found in Nature. Proton and neutron
are the lowest energy states of baryons. Quantum states of baryons are
characterized by means of 3 quarks, called valence quarks. In order to
explain bound states of this kind by means of magnetic monopole forces,
one makes another postulate. Thus, by analogy of atoms, it is assumed that

\begin{itemize}

\item[{(F)}] Baryons have a core whose magnetic charge is $+3g$.

\end{itemize}

This assumption provides an explanation of baryons, using the baryonic
core and 3 quarks attracted to it by the dual Coulomb force which exists
between magnetic monopoles. The sign of the monopole charge of the core
and that of quarks is defined here in order to keep an analogy with the
sign of the electric charge of nuclei and electrons, respectively.

The structure of the baryonic core is beyond the scope of the present
work. It may be an elementary object or, more probably, a complex system
of closed shells of strongly interacting objects whose overall magnetic
charge is $+3g$ (see a remark on this issue in the last Section).

Experiments show that baryons do have a core. Thus, it is found that in an
appropriate Lorentz frame, quarks and antiquarks carry only about one half
of the nucleon's momentum (see [9], p. 282). Hence, something else exists
in a nucleon and here it is called a core.

Postulate (F) leads to further predictions.

\begin{itemize}

\item[{5.}] Like in the case of mesons, bound states of baryons that
cannot be created by 3 quarks are called exotic. Arguments that correspond
to those of point 4 above lead to the claim that {\em strongly} bound
states that make exotic baryons do not exist.
\end{itemize}

This prediction is confirmed experimentally. The lowest bound state of a
baryon and another hadron is the deuteron which consists of one proton and
one neutron. The binding energy of the deuteron is about 2.2 MeV, whereas
gaps between baryonic energy levels are measured by hundreds of MeV. Other
bound states of baryons are nuclei. Here the binding energy is generally
about 8 MeV per nucleon. Another aspect of this point is the similarity
between the form of the nuclear forces and the van-der-Waals ones. This is
the reason for the success of the nuclear liquid drop model[14].

Another issue is the existence of antiparticles in the system. This is a
relativistic effect which may be found more easily in a system where
interaction energies are very high. Thus, experiments show that, beside
the 3 valence quarks, antiquarks exist in nucleons, too. Here antiquarks
are associated with the existence of additional $q\bar {q}$ pairs in the
baryonic wave function (see [9] pp. 282). It turns out that antiquarks
occupy a rather narrow region of the variable $x$ used for characterizing
relations between Lorentz invariant quantities of experiments (see [9], p.
281).

Now, a narrow $x$ pertains to a small Fermi motion in the volume occupied
by the baryon. Thus, due to the uncertainty principle, one concludes that
antiquarks occupy a larger volume than quarks do (see [9], pp. 266-271).

The magnetic monopole theory discussed here explains this point very
easily:

\begin{itemize}

\item[{6.}] If $q\bar q$ pairs are found in a baryonic wave function then
quarks are attracted to the core and antiquarks are pushed away from it.
This result is related to the fact that near the center, the field of the
core's magnetic charge is not completely screened by quarks. Hence,
antiquarks, whose magnetic charge takes the same sign as that of the core,
are pushed away to outer regions.

\end{itemize}

Let us examine the interaction between quarks in a baryon. Like the
interaction between electrons in an atom, this interaction is repulsive
and increases the energy of baryonic states. Hence, if in certain quantum
states, this interaction is reduced then these states should have a higher
binding energy. Since electrons and quarks are spin-1/2 Dirac particles,
their total quantum state is antisymmetric. Now, electronic states whose
total spin is symmetric have a spatial state which is completely
antisymmetric. These states yield a smaller repulsive energy between
electrons than corresponding states where the spatial part is symmetric.
Indeed,  let us examine 2 electrons and ${\bf r}_{12} = {\bf r}_1 -{\bf 
r}_2$ be chosen as a dynamical coordinate. Hence, for ${\bf r}_{12} = 0$, 
an antisymmetric wave function must vanish. This property indicates that
at a region where ${\bf r}_{12}$ is very small and the repulsive
interaction is highest, spatially antisymmetric wave functions yield a
smaller contribution. This effect is related to the Hund rule in atomic
states[15,16]. Antisymmetric spatial wave function of 3 quarks must be
created from 3 different single particle wave functions. Hence, excited
single particle wave functions are used and the system's kinetic energy
increases. Let us have a very rough estimate of this quantity. For this
purpose, consider 2 terms of the state $J^P = 3/2^+$, where the spin and
isospin are symmetric

\begin{equation}
\psi = \psi _1 + \psi _2.
\label{eq:PSI12}
\end{equation}
Here $\psi _1$ is the obvious state $\psi _1 = \phi _1 \phi _2 \phi _3$,
where the $\phi _i$ are single particle S-waves, $\phi _1$ is the ground
state and the other ones are the first and the second radial excitations,
respectively. $\psi _2 = \phi _1 (\chi _1 \chi _2,L=1)$. $\chi _i$ are
single particle P-waves which are coupled to the antisymmetric state[17]
$L=1$. For each of these $\chi _i$, one finds from the spatial angular
momentum

\begin{equation}
1 = l = \mid {\bf r \times p }\mid \Longrightarrow
p_T = \frac {1}{0.8} fm ^{-1} \simeq 250 MeV,
\label{eq:PP}
\end{equation}
where $p_T$ is the momentum component which is perpendicular to $\bf r$
and $0.8fm$ is an estimate of the effective radius. Using the relation
between momentum and kinetic energy, $\Delta P > \Delta E_k$, one finds
from  $(\!\!~\ref{eq:PP})$ that the increase of the kinetic energy of 
$\psi_2$ is about 500 MeV. Here the Hamiltonian is evaluated for $J^P = 
3/2^+$ functions. As is well known, the lowest state obtained after
diagonalization of the Hamiltonian matrix, is lower than the diagonal
entry of each of the basis functions. To this reduction one has to add the
expected contribution of the magnetic monopole analogue of the Hund effect.

Hence,

\begin{itemize}

\item[{7.}]  The mass of the $\Delta _{1232}$ baryon, which is higher than
the nucleonic mass by about 300 MeV, is understood.

\end{itemize}

Conclusion 1 and the related equations of motion
$(\!\!~\ref{eq:FINALLORE})$ and $(\!\!~\ref{eq:FINALLORM})$ indicate that
static electric field of a charge and electric field of a moving monopole
have different dynamical properties. The same conclusion holds for the
corresponding magnetic fields. A special case of this distinction is found
in the electric field of a polar dipole (which is made of two displaced
electric charges having equal strength and opposite sign) and that of an
axial electric dipole of a spinning monopole. The {\em axial} electric
dipole of spinning monopoles is discussed here.

The neutron is known to be a spin-1/2 electrically neutral composite
particle. Its nonvanishing magnetic dipole moment demonstrates that not
all effects of its electrically charged constituents vanish. The duality
relations between electric charges and magnetic monopoles provide the
basis for the following statement. If quarks are dyons (namely, particles
that have both electric and magnetic charge) and strong interactions are
interactions between magnetic monopoles then, it is highly reasonable that
neutrons (and protons) should have a large {\em axial} electric dipole
moment which is associated with spinning monopoles. Indeed, it is
extremely unlikely that the overall electric dipole moment of a system of
spinning monopoles vanishes whereas the total spin is nonzero. This
discussion indicates that the very low upper bound measured for the
electric dipole moment of the neutron[18,19] should not be regarded as a
major argument against a hadronic theory where quarks are magnetic
monopoles obeying $(\!\!~\ref{eq:FINALLORE})$ and
$(\!\!~\ref{eq:FINALLORM})$. Indeed, all experiments carried out for the
measurement of the electric dipole moment of the neutron are eventually
based on the interaction of electric field of charge with the searched
electric dipole moment of the neutron[18,19].

Thus, the very low upper bound measured for the electric dipole moment of
neutrons is, as a matter of fact, an upper bound for its {\em polar}
electric dipole moment. These measurements provide no information on the
magnitude of the neutron's {\em axial} electric dipole moment. Hence,
results of measurements of the neutron's electric dipole moment are not
incompatible with the regular charge-monopole theory presented in this
work whose main results are $(\!\!~\ref{eq:FINALLORE})$ and
$(\!\!~\ref{eq:FINALLORM})$.

As pointed out above, a nucleon is expected to have a nonvanishing {\em
axial} electric dipole moment, due to its spinning quarks. In this way,
one finds an explanation for the tensor interaction between nucleons[20,21]

\begin{equation}
V_T = \{3(\mbox{\boldmath $\sigma $}_1 \mbox{\boldmath $\cdot r$})
(\mbox{\boldmath $\sigma$}_2 \mbox{\boldmath $\cdot r$})
-r^2\mbox{\boldmath $\sigma $}_1\mbox{\boldmath $\cdot \sigma $}_2\}
U(r),
\label{eq:TENSOR}
\end{equation}
where {\boldmath $r$}={\boldmath $r$}$_2$-{\boldmath $r$}$_1$ and
{\boldmath $\sigma $} is the spin operator. This expression is a
generalization of the dipole-dipole interaction between two static point
dipoles {\boldmath $\mu $}$_1$ and {\boldmath $\mu $}$_2$ (see [2], p
143).

\begin{equation}
V_{DIPOLE} = -\{3(\mbox{\boldmath $\mu $}_1
\mbox{\boldmath $\cdot r$})
(\mbox{\boldmath $\mu $}_2 \mbox{\boldmath $\cdot r$})
-r^2\mbox{\boldmath $\mu $}_1\mbox{\boldmath $\cdot \mu $}_2\}
/r^5.
\label{eq:DIPDIP}
\end{equation}

Evidently, the nuclear tensor interaction cannot be exactly a
dipole-dipole one, because nucleons are not point dipoles but composite
particles whose size is not much smaller than the distance between
nucleons in a nucleus. For this reason the form of the function $U(r)$ of
$(\!\!~\ref{eq:TENSOR})$ is determined phenomenologically. It is
interesting to note that the {\em sign} of $U(r)$ of
$(\!\!~\ref{eq:TENSOR})$ is negative (see [21], p. 103), like the sign of
$(\!\!~\ref{eq:DIPDIP})$. It can be concluded that

\begin{itemize}

\item[{8.}] The origin of the nuclear tensor force is understood.

\end{itemize}

The size of the nucleonic volume inside a nucleus can be deduced from the
nucleonic quarks' momentum. It is found that the larger the number of
nucleons $A$ in a nucleus, the larger is the mean self volume of nucleons
of this nucleus. In other words, as the nucleus becomes heavier its
nucleons swell. This property is compatible with the EMC effect[22,23].

On the other hand, the success of the nuclear liquid drop models is an
indication that nuclear density is practically constant (see [14] and
[20], pp. 6,7). The swelling of the mean volume occupied by quarks of a 
nucleon with the increase of the number $A$ of nucleons in nuclei, is 
compatible with screening properties of electrodynamics. Consider a 
nucleon $N_i$ in a nucleus. A part of the wave function of quarks of 
neighboring nucleons penetrates into the volume occupied by $N_i$. Thus, 
the attracting field of the core of $N_i$ is partially screened by quarks 
belonging to neighboring nucleons. It follows that, in this case, quarks 
of $N_i$ ``see" a weaker field attracting them to the core of $N_i$ and 
settle in a larger volume. As the number of nucleons of a nucleus, $A$, 
increases, the average number of neighbors of a typical nucleon increases 
too and the screening effect becomes more significant. This situation 
explains the EMC effect. Thus,

\begin{itemize}

\item[{9.}] screening effects cause self volume of quarks of each nucleon
to increase inside rather large nuclei.

\end{itemize}

The 3 valence quarks of the proton are $uud$. Thus, one may write a
truncated sum of terms of the proton's full wave function as follows:

\begin{equation}
\psi _{proton} = \lambda _0 \phi_0 (uud)  + \lambda _1 \phi_1 (uudu\bar
{u}) + \lambda _2 \phi_2 (uudd\bar {d}).
\label{eq:PROTONWF}
\end{equation}
Here  $\phi _0$ of $(\!\!~\ref{eq:PROTONWF})$ denotes a wave function of
the 3 valence quarks (and the completely full ``sea" of negative energy
states of quarks). In $\phi _1$, one $u$ quark is excited from the ``sea"
into a positive energy state. $\phi _2$ is analogous. Every $\phi_i$ is
normalized and each of the $\lambda _i$ is a numerical coefficients.
Obviously, each of these terms has the proton's quantum numbers.
Since the valence quarks of the proton contain 2 $u$ quarks and only one
$d$ quark, it is obvious that the additional $d$ quark of $\phi _2$ finds
a lower energy state than the additional $u$ quark of $\phi _1$. Hence,
the absolute value of the coefficient $\lambda _2$ should be greater than
that of $\lambda _1$. It can be concluded that

\begin{itemize}

\item[{10.}] the regular charge-monopole theory provides an explanation
for the extra $\bar {d}$ found in a proton, called flavor asymmetry[24].

\end{itemize}

The 10 issues pointed out in this Section should be considered as an
illustration of the ability of the regular charge-monopole theory to
explain phenomena related to strong interaction.

\section{A Comparison of Magnetic Monopole Theories}

This Section contains a discussion of the Dirac monopole theory and its
comparison with the regular charge-monopole theory outlined in Section 2.
An introductory part is needed for clarifying some general aspects. A
physical theory is a mathematical structure that has a physical domain of
validity [25]. Hence, in principle, a theory can be refuted if its
mathematical structure leads to a contradiction. Its physical meaning can
be rejected if it fails to explain results of physical measurements
carried out within its domain of validity. It should be pointed out that
in the latter case, the theory cannot be saved by an attempt to
gerrymander its validity domain[25]. The foregoing arguments indicate that
one cannot refute a theory by means of another theory. Indeed, assume that
theories A and B yield contradictory predictions within a common validity
domain. In this case it may be concluded that at least one of these
theories is wrong but it is still unknown which theory is the wrong one.
Thus, a comparison with experimental data is required.

There is another aspect of a theory which is not directly connected to its
correctness and has also a subjective personal side. It is generally
accustomed to regard a theory as a neat one if it relies on a minimal set
of postulates which are based on general properties of Nature. One also
generally expects that a neat theory is more likely to be correct when
compared to another theory which does not look neat. For this reason, the
specific postulates used by theories are mentioned too.

The origin of problems of the Dirac charge-monopole theory is that it does
not start with the simple case of a system of monopoles without charges.
As shown in Section 2, an examination of this case together with postulate
(A) leads to the regular charge-monopole theory where the Lorentz force
takes the form of $(\!\!~\ref{eq:FINALLORE})$ and
$(\!\!~\ref{eq:FINALLORM})$. Let us turn to some problematic points of the
Dirac charge-monopole theory. These points show that Dirac's
charge-monopole theory differs drastically from electrodynamics.

\begin{itemize}

\item[{1.}] The Dirac theory yields a kind of irregularity which is not
found in other parts of classical electrodynamics. Indeed, let us examine
a magnetic charge density $\rho _m$ and the Dirac's vector potential,
which is used for fields of charges {\em and} for fields of monopoles

\begin{equation}
4\pi \rho _m = \nabla \cdot {\bf B} = \nabla \cdot (\nabla \times {\bf A})
= 0.
\label{eq:DIVB}
\end{equation}

This relation states that monopoles do not exist if the vector potential
${\bf A}$ is regular. Hence, in Dirac's monopole theory the vector
potential must be singular. For this reason, Dirac has introduced the
notion of a string that extends from a monopole $g$ to infinity or ends at
another monopole having a magnetic charge $-g$. Moreover, the system
should obey what is called Dirac's veto which forbids charges from
entering regions of space occupied by strings (see [6], p. 1374). Strings
are a {\em new} notion introduced into the theory. Thus, the following
postulate enables their acceptability.

\item[{(A')}] Strings connected to monopoles account for the
irregularities of the theory. Charges are not allowed to enter regions of
space occupied by strings.

\item[{2.}] Unlike ordinary classical electrodynamics and the regular
charge-monopole theory outlined in Section 2, Dirac's charge-monopole
theory cannot be derived from a regular Lagrangian[4,26]. Hence, the
definition of canonical variables is unclear and the ordinary method of
constructing the Hamiltonian cannot be used. For this reason, the form of
quantum mechanics of Dirac's charge-monopole theory is not self-evident.
Analogous conclusions have already been published[27-29].

\end{itemize}

Let us examine the angular momentum of a Dirac charge-monopole system. A
ring made of an insulating material is placed in the $(x,y)$ plane and its
center coincides with the origin. The ring is covered uniformly with
charge density $\rho $ and it can rotate around the $z$-axis. A monopole
$g$ moves along the $z$-axis from $-\infty $ towards $\infty $ (see fig.
1) and carries its string along its path. The motion is legitimate
according to Dirac's veto, because charges of the ring do not touch the
$z$-axis. Now, the motion of the monopole is accompanied with a circular
electric field which is dual to the circular magnetic field of a uniformly
moving charge. Due to Dirac's monopole theory, this field accelerates
charges along the ring. Hence, the value of the angular momentum of the
ring at the final time differs from its value at the initial time. It
turns out that this variation of the angular momentum is compensated by
the interaction part of a charge-monopole system (see [2], p. 256 and [6],
pp. 1365, 1366. Note the different units used in [2] and [6]). Thus, in
Dirac's monopole theory, the angular momentum of a static system of a
charge and a monopole is

\begin{figure}[b]
\begin{center}
\epsfig{file=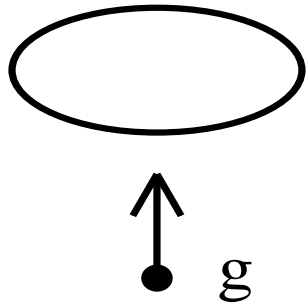,width=4cm,height=4cm}
\caption{A monopole $g$ moves through a ring (see text).}
\label{fig:1}
\end{center}
\end{figure}

\begin{equation}
{\bf j} = eg {\bf r}/r
\label{eq:JDIRAC}
\end{equation}
where $\bf r$ denotes the radius vector from the charge to the monopole.
This angular momentum depends on the {\em direction} of the line
connecting the charge and the monopole but is {\em independent of the
distance between the particles}. It follows that

\begin{itemize}

\item[{3.}] in Dirac's charge-monopole theory there is an interaction
dependent quantity whose value does not vanish even if the particles are
infinitely far apart. This result is strange and unconvincing.

\item[{4.}] Eq. $(\!\!~\ref{eq:JDIRAC})$ as well as other arguments yield
the results that in Dirac's monopole theory, a magnetic charge $g$
transforms not like a scalar but like a pseudoscalar (see [2], p. 253). It
is not clear how this outcome affects the theory. In particular, by
analogy with charges (see [1], p. 70, [2], p. 549), one expects that the
monopole 4-current is $\rho _m v^\mu/\gamma $, where $\rho _m$ is the
monopole density and $\gamma = (1-v^2)^{-1/2}$. Now, one expects that the
4-current behaves like a 4-vector, contrary to the pseudoscalar property
assigned to monopoles.

\end{itemize}

Assume that, in spite of what is said in item 2 of this section, one finds
a way for introducing monopoles into quantum mechanics. In this case, the
self consistency of the theory is doubtful because the theory is not
derived from a regular Lagrangian. The following example illustrates this
issue. Let us examine a spin 1/2 charged particle which is attracted to a
center of force by a non-electromagnetic interaction (henceforth denoted
by NEMI). NEMI is much stronger than all electromagnetic forces and the
latter are evaluated by perturbation calculations. A peculiar feature of
NEMI is that its spin-orbit interaction is very strong and its $l=3$
$j=7/2$ state is its ground state. An external field of the NEMI performs
a split of the $m$ states, which is analogous to the Zeeman effect. Thus,
the state $l=3$, $j=7/2$, $m_j =7/2$ is the ground state. This is a
quantum mechanical analogue of the ring of fig. 1. A magnetic monopole $g$
moves along the $z$-axis from $z=-Z_0$ to $z=Z_0$ and returns back along a
semicircle whose center is at the origin ($Z_0$ is very large). Evidently,
the final state equals the initial one, except for the energy gained by
the monopole while moving along the $z$-axis. (Energy involved with the
motion along the arc can be ignored because on this line the magnetic
field of the magnetic dipole of the $j=7/2$ charge behaves like $r^{-3}$.)
Hence,

\begin{itemize}

\item[{5.}]  quantum mechanics of the Dirac monopole theory does not
conserve energy.

\item[{6.}] Another aspect related to the Dirac monopole theory is the
usage of measurement of fields by charges and by monopoles. It can be
shown that if the laboratory is located in a noninertial frame of
reference (like a rotating laboratory) then fields measured by charges and
fields measured by monopoles are {\em different} entities [30]. Hence, it
is not clear why, in Dirac's charge-monopole theory, these different
entities are derived from the same 4-potential.

\item[{7.}] Beside the foregoing theoretical difficulties, there is the
experimental situation stating that, in spite of prolonged efforts, Dirac
monopoles have not been detected[13,31]. This outcome has been predicted
on the basis of S-matrix considerations[27] and by using the regular
charge-monopole theory outlined in Section 2[32]. This is probably the
only monopole related prediction that still holds till now.

\item[{8.}] It should be mentioned that, in addition to the problematic
points of the Dirac monopole theory, this theory explains a basic
phenomenon of Nature which is charge quantization. However, Dirac's value
of the elementary monopole unit $g^2 \simeq 34$ appears to be very high.

\end{itemize}

Charge quantization clearly cannot be derived from the regular monopole
theory of Section 2. However, it can be argued that it is very far from
being self-evident that a proof of charge quantization is an inseparable part of the
validity domain of an acceptable charge-monopole theory.

\section{Problems of Quantum Chrodynamics}

At present, quantum chrodynamics (QCD) is regarded as the theory of strong
interactions (see [9], p. 20). It relies on several postulates that have
not been used previously. By analogy with the discussion in Section 3,
these postulates are described below.

\begin{itemize}

\item[{A''.}] It uses the Young-Mills idea which extends the gauge
procedure  and enables the replacement of complex numbers of the phase by
matrices belonging to a certain group. In the case of QCD, the group is
$SU_3$.

\item[{B''.}] It relies on a certain procedure called after Higgs, which
enables the theory to take an acceptable form. The Higgs procedure
requires the existence of particles called Higgs mesons.

\item[{C''.}] It assigns to each quark a new kind of charge called color.
There are 3 kinds of color, called red, green and blue, each of which has
its own anticolor. The theory further assumes that particles found in
laboratories must be white, namely, they should have equal amount of
positive and negative values of each color or equal amount of all the 3
colors.

\end{itemize}

Let us make a list of problematic points of QCD. The list has a certain
resemblance to some of the points discussed in Section 3, where the
utilization of the regular charge-monopole theory to strong interactions
is described.

\begin{itemize}

\item[{1.}] In spite of a long experimental search, the Higgs mesons have
not been found [13].

\item[{2.}] QCD provides no explanation to the participation of real
photons in strong interactions. It should be pointed out that the approach
called Vector Meson Dominance as well as similar ideas have been discussed
recently. It is proved that these approaches have no theoretical basis[33].

\item[{3.}] QCD does not rule out the existence of exotic states (see,
e.g. [13] pp. 754, 755, [34]). However, the validity of these states is
not established by experiments.

\item[{4.}] The fact that quarks carry only about one half of the
nucleon's momentum is explained in Section 3 by the introduction of the
baryonic core (see the discussion that follows postulate (F)). In QCD,
this is explained by the claim that gluons (the QCD analogue of
electromagnetic bound fields) carry the rest of the momentum. These
different explanations can be tested by an analysis of sufficient data
obtained from colliding beams of $\pi $ mesons and electrons. Here, the
monopole theory used in Section 3 predicts that quarks of a $\pi $ meson
should carry all momentum whereas QCD expects that gluons of the $q\bar q$
pair of the $\pi$ meson should take a portion of the momentum.

\end{itemize}

\clearpage

\section{Concluding Remarks}

This work shows how a regular charge-monopole theory can be derived from
the self-evident duality postulate. It is also shown that this theory can
be applied to the field of strongly interacting particles. This
application relies on very simple and self-evident postulates and explains
several qualitative properties of strong and nuclear interaction. Thus, it
removes the asymmetry between electricity and magnetism in contemporary
electrodynamics.

The idea that baryons contain a magnetic charge has already been
published. Even before the discovery of quarks, it has been suggested by
Dirac that nucleon constituents contain monopoles [8]. Schwinger has
proposed a model of hadrons where quarks are dyons[35,36]. This course has
been examined by other authors, too [37,38]. However, since all these
authors have used the Dirac charge-monopole theory, they could not
overcome difficulties. For example, the Dirac monopole theory cannot
explain why electrons do not ``see" the monopole field but real photons do
``see" it (see item 2 of Section 3). Similarly, it cannot explain why the
axial electric dipole moment associated with spinning monopoles is not
detected in neutron measurements (see the discussion that follows item 7
of section 3).

Problems and difficulties of the Dirac monopole theory are discussed in
Section 4. These difficulties and the failure of the experimental quest
for Dirac monopoles, indicate that it is unlikely that the Dirac
charge-monopole theory is correct.
Several kinds of experimental data which are not explained by QCD are
discussed in Section 5. The data certainly belongs to the domain of
validity of QCD or to the wider theory called the Standard Model. Thus,
the interaction of a real photon with a hadron belongs to the combined
domain of electrodynamics and QCD, both of which are elements of the
Standard Model. As proved elsewhere [33], QCD does not provide an
acceptable explanation for this phenomenon.

Other points, like the failure to detect the Higgs mesons as well as
exotic states of hadrons belong to the validity domain of QCD. Moreover,
QCD has not predicted the EMC effect [22]. It is shown in Section 3 how
easily the monopole theory of hadrons explains the similarity between the
van der Waals forces and the nuclear ones, as well as the nuclear tensor
force, including its sign. On the other hand, in spite of intensive work
carried out during more than 30 years, standard textbooks on QCD[9,10] do
not discuss these topics.

Similarly, if the baryonic core is assumed to consist of a magnetic
monopole central object and closed shells of inner quarks then the
enhancement found in the cross section of very high energy
collisions[39,40] is understood. These experiments indicate the existence
of an energy threshold above which additional quarks at the nucleonic
target begin to participate in the interaction with the projectile. The
analysis of deep inelastic scattering of electrons on $\pi $ mesons may
provide new evidence concerning the different interpretations of the
portion of momentum carried by nucleonic quarks. In QCD the rest of the
momentum is ascribed to gluons whereas in the monopole theory outlined
here, it is ascribed to the baryonic core that carries also 3 units of
monopole charge. In the case of mesons, there is no core and the monopole
theory predicts that all momentum is carried by quarks and antiquarks. On
the other hand, according to the QCD approach, a portion of the $\pi $
meson's momentum is expected to be carried by gluons.

A theory differs from a model by the fact that a theory is expected to fit
accurately results of experiments carried out within its validity domain.
For this reason, it can be concluded that QCD's inability to describe
correctly well established experimental data casts doubts on its
correctness.

\nonumsection{References}

\end{document}